\documentstyle[preprint,eqsecnum,aps]{revtex}
\begin{document}
\title{One-Loop Effective Action for Euclidean Maxwell Theory\\
on Manifolds with Boundary}
\author{Giampiero Esposito,$^{1,2}$
\thanks{Electronic address: esposito@napoli.infn.it}
Alexander Yu. Kamenshchik$^{3}$
\thanks{Electronic address: grg@ibrae.msk.su}
and Klaus Kirsten$^{4}$
\thanks{Electronic address: kirsten@tph100.physik.uni-leipzig.de}}
\address{${ }^{1}$Istituto Nazionale di 
Fisica Nucleare, Sezione di Napoli,\\
Mostra d'Oltremare Padiglione 20, 80125 Napoli, Italy\\
${ }^{2}$Dipartimento di Scienze Fisiche,\\
Mostra d'Oltremare Padiglione 19, 80125 Napoli, Italy\\
${ }^{3}$Nuclear Safety Institute, Russian Academy of Sciences,\\
52 Bolshaya Tulskaya, Moscow 113191, Russia\\
${ }^{4}$Universit\"{a}t Leipzig, 
Institut f\"{u}r Theoretische Physik,\\
Augustusplatz 10, 04109 Leipzig, Germany}
\maketitle
\begin{abstract}
This paper studies the one-loop effective action for Euclidean
Maxwell theory about flat four-space bounded by one three-sphere,
or two concentric three-spheres. The analysis relies on
Faddeev-Popov formalism and $\zeta$-function regularization,
and the Lorentz gauge-averaging term is used with magnetic
boundary conditions. The contributions of transverse, longitudinal
and normal modes of the electromagnetic potential, jointly with
ghost modes, are derived in detail. The most difficult part of 
the analysis consists in the eigenvalue condition given by the
determinant of a $2 \times 2$ or $4 \times 4$ matrix for
longitudinal and normal modes. It is shown that the former splits
into a sum of Dirichlet and Robin contributions, plus a simpler
term. This is the quantum cosmological case. In the latter case,
however, when magnetic boundary conditions are imposed on two
bounding three-spheres, the determinant is more involved. 
Nevertheless, it is evaluated explicitly as well. The whole analysis
provides the building block for studying the one-loop effective 
action in covariant gauges, on manifolds with boundary. The final
result differs from the value obtained when only transverse modes
are quantized, or when noncovariant gauges are used.
\end{abstract}
\section{Introduction}

Over the last few years, a considerable effort has been 
produced in the literature to study the problem of boundary
conditions in Euclidean quantum gravity and quantum cosmology
[1--10]. This is motivated by the need to obtain a well defined
path-integral representation of the $\langle {\mbox {out}} \mid
{\mbox {in}} \rangle$ amplitudes and of the quantum state of the
universe, and lies at the very heart of any attempt to
understand the basic features of a theory of the quantized
gravitational field. In particular, within the framework of the
semiclassical approximation for the wave function of the
universe, this analysis has led to the first calculation of
one-loop divergences for massless spin-${1\over 2}$ fields
[11--16], Euclidean Maxwell theory [17--20], supergravity models [21]
and Euclidean quantum gravity [6--10] in the presence of 
boundaries. Focusing on massless models, flat Euclidean 
backgrounds bounded by a three-sphere have been studied in detail
for fields of various spins, with local or nonlocal boundary
conditions, and in covariant as well as noncovariant gauges for
gauge theories and gravitation [6--21]. For these fields the boundary
conditions are mixed in that some components of the field obey
one set of boundary conditions, and the remaining components 
obey another set of boundary conditions. This is indeed necessary
to ensure invariance of the whole set of boundary conditions 
under infinitesimal gauge transformations, as well as their 
BRST invariance. For example, one may consider Euclidean Maxwell
theory (which is the object of our investigation). In the
classical theory, one may begin by fixing at the boundary the
tangential components $A_{k}$ of the electromagnetic potential. 
Such a boundary condition is invariant under infinitesimal gauge
transformations ${ }^{\xi}A_{k}=A_{k}+\partial_{k}\xi$ if and
only if $\xi$ itself vanishes at the boundary. In the semiclassical
approximation of quantum theory, one expands about a vanishing
background value for $A_{k}$, so that the electromagnetic potential
reduces to pure perturbations ${\cal A}_{k}$, say. Moreover, if
one follows the Faddeev-Popov method, one adds a gauge-averaging
term ${1\over 2\alpha}[\Phi({\cal A})]^{2}$ to the original Lagrangian,
jointly with a ghost term which is necessary to ensure gauge
invariance of the quantum theory. The first set of boundary
conditions are now
\begin{equation}
\Bigr[{\cal A}_{k}\Bigr]_{\partial M}=0 ,
\end{equation}
\begin{equation}
[\varphi]_{\partial M}=0 ,
\end{equation}
where $\varphi$ is a complex-valued ghost zero-form, 
corresponding to two {\it independent, real} ghost fields
[22], which are both subject to homogeneous Dirichlet
conditions at the boundary. At this stage, the only choice 
of boundary conditions on the normal component ${\cal A}_{0}$,
whose gauge invariance is again guaranteed by the imposition of
(1.2) on the ghost zero-form, is
\begin{equation}
\Bigr[\Phi({\cal A})\Bigr]_{\partial M}=0 .
\end{equation}
If the Lorentz gauge-averaging functional is chosen, Eqs. (1.1)
and (1.3) lead to Robin conditions on ${\cal A}_{0}$, i.e.,
\begin{equation}
\left[{\partial {\cal A}_{0}\over \partial \tau}
+{\cal A}_{0}{\mbox {Tr}}K\right]_{\partial M}=0 ,
\end{equation}
where $K$ is the extrinsic-curvature tensor of the boundary. 

The analysis of conformal anomalies and one-loop divergences,
however, is part of a more general program devoted to the
investigation of the one-loop effective action on manifolds
with boundary. As shown in Refs. [23--26], complete results are
by now available for scalar and spin-${1\over 2}$ fields. For
Euclidean Maxwell theory, the contribution of transverse modes
was obtained in Refs. [25,26], and the contribution of all
perturbative modes in a noncovariant gauge was first obtained
in Ref. [27]. For supergravity and quantum gravity, the
contribution of transverse-traceless modes only has been 
obtained in Refs. [25,26]. It has been therefore our aim to
present a detailed calculation of one-loop effective action in
a covariant gauge including all perturbative modes of the
problem in the presence of boundaries. This is necessary both
for the sake of completeness, and to check whether the contributions
of longitudinal, normal and ghost modes cancel each other exactly
on such bounded regions. No such a cancellation was indeed found
to occur in the analysis of conformal anomalies in Refs. [18--20].

Since the building block of our investigation is the study of
real, massless scalar fields on the four-ball, subject to
Dirichlet conditions, we find it helpful to present here a
very brief outline of such a calculation. As shown in Refs.
[23,28], the starting point is the integral representation of the
$\zeta$-function of a self-adjoint, positive-definite elliptic 
operator by means of the Cauchy formula, which makes it possible
to express $\zeta(s)$ of the Laplace operator in four dimensions
as the sum
\begin{equation}
\zeta(s)=\sum_{l=1}^{\infty}l^{2}Z_{l}(s)
+\sum_{i=-1}^{3}A_{i}(s) .
\end{equation}
By using the notation in the Appendix for the uniform asymptotic
expansions of Bessel functions, one has in (1.5)
\begin{equation}
Z_{l}(s)= {\sin(\pi s)\over \pi}\int_{0}^{\infty}
dz (zl/a)^{-2s}{\partial \over \partial z} \Bigr[
\ln I_{l}(lz)- l \eta 
+ \ln \biggr(\sqrt{2\pi l}(1+z^{2})^{1\over 4}\biggr)
-\sum_{k=1}^{3}{D_{k}(t)\over l^{k}} \biggr] ,
\end{equation}
\begin{equation}
A_{-1}(s)={1\over 4\sqrt{\pi}}{a^{2s}\over \Gamma(s)}
{\Gamma(s-{1\over 2})\over s}\zeta_{R}(2s-3) ,
\end{equation}
\begin{equation}
A_{0}(s)=-{1\over 4}a^{2s}\zeta_{R}(2s-2) ,
\end{equation}
\begin{equation}
A_{1}(s)=-{a^{2s}\over \Gamma(s)}\zeta_{R}(2s-1)
\sum_{j=0}^{1}x_{1,j}{\Gamma(s+j+{1\over 2})\over
\Gamma(j+{1\over 2})} ,
\end{equation}
\begin{equation}
A_{2}(s)=-{a^{2s}\over \Gamma(s)}\zeta_{R}(2s)
\sum_{j=0}^{2}x_{2,j}{\Gamma(s+j+1)\over
\Gamma(j+1)} ,
\end{equation}
\begin{equation}
A_{3}(s)=-{a^{2s}\over \Gamma(s)}\zeta_{R}(2s+1)
\sum_{j=0}^{3}x_{3,j}{\Gamma(s+j+{3\over 2})\over
\Gamma(j+{3\over 2})} ,
\end{equation}
where $a$ is the radius of the three-sphere boundary. Thus,
the algorithms described in Refs. [23,27] lead to the following
result for the one-loop effective action:
\begin{eqnarray}
\Gamma^{(1)}&=&-{1\over 2}\zeta'(0)
-{1\over 2}\zeta(0)\ln(\mu^{2}) \nonumber\\
&=&{1\over 360}\ln(\mu^{2}a^{2})-{1\over 180}\ln(2)
-{173\over 60480}-{1\over 6}\zeta_{R}'(-3)
+{1\over 4}\zeta_{R}'(-2)-{1\over 12}\zeta_{R}'(-1) .
\end{eqnarray}

Section II evaluates the one-loop effective action for Euclidean
Maxwell theory in a background motivated by quantum cosmology,
i.e., flat Euclidean four-space bounded by a three-sphere. This
results from the analysis of the wave function of the universe
in the limit of small three-geometries [29]. Magnetic boundary
conditions in the Lorentz gauge, i.e., (1.1), (1.2) and (1.4), are
imposed. Section III studies instead the occurrence of two
concentric three-sphere boundaries. This case is more relevant for
quantum field theory. Section IV presents some independent
derivations, based on the technique developed by Barvinsky,
Kamenshchik, Karmazin and Mishakov.
Results and open problems are discussed
in Sec. V, and relevant details are given in the Appendix.  

\section{One-boundary problem}

In this section we study a background four-geometry given by
flat Euclidean four-space bounded by a three-sphere. Since the
boundary three-geometry is $S^{3}$, this ensures that the
tangential components of the electromagnetic perturbations 
consist of a transverse part, ${\cal A}_{k}^{T}$, and a
longitudinal part, ${\cal A}_{k}^{L}$, only [30]. These are
expanded on a family of three-spheres centered on the origin
as [31]
\begin{equation}
{\cal A}_{k}^{T}(x,\tau)=\sum_{n=2}^{\infty}f_{n}(\tau)
S_{k}^{(n)}(x) ,
\end{equation}
\begin{equation}
{\cal A}_{k}^{L}(x,\tau)=\sum_{n=2}^{\infty}g_{n}(\tau)
P_{k}^{(n)}(x) ,
\end{equation}
where $\tau$ is a radial coordinate $\in [0,a]$, $x$ are local
coordinates on $S^{3}$, and $S_{k}^{(n)}$ and $P_{k}^{(n)}$ are
transverse and longitudinal vector harmonics on the three-sphere,
respectively [32]. Moreover, the occurrence of the boundary with
normal vector $n^{\mu}$ makes it possible to define the normal
component of $A_{\mu}(x,\tau)$ as $A_{0}(x,\tau) \equiv
n^{\mu}A_{\mu}(x,\tau)$. Its expansion on the same family of
three-spheres reads [31]
\begin{equation}
{\cal A}_{0}(x,\tau)=\sum_{n=1}^{\infty}R_{n}(\tau)
Q^{(n)}(x) ,
\end{equation}
where $Q^{(n)}$ are scalar harmonics on $S^{3}$ [32]. We are
actually facing a crucial point in our analysis, since the
singularity at the origin of our background four-manifold calls
into question the validity of the 3+1 split (2.1)--(2.3) 
{\it inside} the bounding three-sphere of radius $a$ [18--20]. 
In the analytic approach, one requires that the expansions
(2.1)--(2.3) should match the boundary values 
${\cal A}_{k}^{T}(x,a),{\cal A}_{k}^{L}(x,a),
{\cal A}_{0}(x,a)$, and be regular $\forall \tau \in [0,a]$.
This should pick out a unique smooth solution [17]. In the
geometric analysis it seems that, as long as the operator acting
on ${\cal A}^{\mu}$ reduces to $-g_{\mu \nu}\Box$ in flat four-space,
the analysis of the problem remains well defined at any stage,
since it does not contain any (explicit) reference to ill-defined
objects or noncovariant elements (e.g., ${\mbox {Tr}}K$ terms in
the differential operators). Indeed, in covariant gauges, the
operator on ${\cal A}^{\mu}$ is
$$
-g_{\mu \nu}\Box +R_{\mu \nu}+\Bigr(1-{1\over \alpha}\Bigr)
\nabla_{\mu}\nabla_{\nu} ,
$$
and this reduces to the desired $-g_{\mu \nu}\Box$ in flat
four-space, provided that one makes the Feynman choice for the
$\alpha$ parameter: $\alpha_{F}=1$.

The contribution of transverse modes $f_{n}(\tau)$ in (2.1) is
independent of any choice of gauge-averaging term in the
Faddeev-Popov Euclidean action, since such modes are decoupled
from longitudinal modes ($g_{n}$) and normal modes ($R_{n}$).
Thus, relying on the $\zeta_{T}'(0)$ value obtained in Refs.
[25,26], one finds
\begin{equation}
\Gamma_{T}^{(1)}={77\over 360}\ln(\mu^{2}a^{2})
+{29\over 90}\ln(2)+{1\over 2}\ln(\pi)+{6127\over 30240}
-{1\over 3}\zeta_{R}'(-3)+{1\over 2}\zeta_{R}'(-2)
+{5\over 6}\zeta_{R}'(-1) .
\end{equation}

It is also straightforward to obtain the contribution of
ghost modes. Bearing in mind that they behave as fermionic
modes (if the gauge field is bosonic, as in our case), and
imposing the boundary condition (1.2), one has simply to 
multiply by $-2$ the scalar-field result (1.12). This leads
to
\begin{equation}
\Gamma_{\mbox {ghost}}^{(1)}=-{1\over 180}\ln(\mu^{2}a^{2})
+{1\over 90}\ln(2) +{173\over 30240}+{1\over 3}\zeta_{R}'(-3)
-{1\over 2}\zeta_{R}'(-2)+{1\over 6}\zeta_{R}'(-1) .
\end{equation}

The only technical difficulties consist in the analysis
of coupled longitudinal and normal modes. As shown in Ref. [19],
the boundary conditions (1.1) and (1.4) lead to an eigenvalue
condition for such modes given by the vanishing of the determinant
of a $2 \times 2$ matrix. With the notation of Ref. [19], in the
Lorentz gauge this equation reads ($\forall n \geq 2$)
\begin{eqnarray}
\;&\;& I_{n+1}(Ma) \Bigr(2{I_{n-1}(Ma)\over (Ma/2)}+I_{n-2}(Ma)
+I_{n}(Ma)\Bigr) \nonumber\\
&+& {(n+1)\over (n-1)}I_{n-1}(Ma)\Bigr(2{I_{n+1}(Ma)\over
(Ma/2)}+I_{n}(Ma)+I_{n+2}(Ma)\Bigr)=0 .
\end{eqnarray}
Thus, the first step is to re-express 
$I_{n-2},I_{n-1},I_{n+1}$ and $I_{n+2}$ in terms of $I_{n}$
and $I_{n}'$ only. By virtue of Eqs. (A16)--(A19) of the
Appendix, and setting $Ma=zn$, Eq. (2.6) is found to involve 
the following function of $z,I_{n}$ and $I_{n}'$:
\begin{equation}
{\cal F}={1\over z}I_{n}(zn)\Bigr(I_{n}(zn)
+znI_{n}'(zn)\Bigr) ,
\end{equation}
where proportionality parameters have been omitted, since they
do not affect the calculation of $\zeta'(0)$. The function (2.7)
should be inserted into the integral representation of the
$\zeta$-function for longitudinal (L) and normal (N) modes,
i.e.,
\begin{equation}
\zeta_{LN}(s)={\sin(\pi s)\over \pi}\sum_{n=2}^{\infty}
n^{2} \int_{0}^{\infty}dz (zn/a)^{-2s}
{\partial \over \partial z}
\ln \Bigr[z^{-2n+1}{\cal F}\Bigr] ,
\end{equation}
where $n^{2}$ is the degeneracy of such modes. Remarkably,
after re-expressing the infinite sum in (2.8) as an infinite
sum from $1$ to $\infty$, minus the contribution of $n=1$,
one finds, by virtue of (2.7), that the resulting contribution
to $\zeta'(0)$ is the sum of three contributions: the 
$\zeta'(0)$ value for a real scalar field subject to Dirichlet 
conditions on $S^{3}$; the $\zeta'(0)$ value for a real
scalar field subject to Robin conditions on $S^{3}$, with $u$
parameter equal to 1; the $\zeta'(0)$ value resulting from 
$n=1$. The first is given in Ref. [23] and is encoded in the
one-loop effective action (1.12). The second is evaluated in
Ref. [33] as
\begin{equation}
\zeta_{{\mbox {Robin}},u=1}'(0)=-{41\over 864}
-{7\over 45}\ln(2)-{1\over 2}\ln(\pi)-{31\over 90}\ln(a)
+{1\over 3}\zeta_{R}'(-3)+{1\over 2}\zeta_{R}'(-2)
-{11\over 6}\zeta_{R}'(-1) .
\end{equation}
The third is obtained from (see (2.7) and (2.8))
\begin{equation}
{\widetilde \zeta}(s) \equiv -{\sin(\pi s)\over \pi}
\int_{0}^{\infty}dz (z/a)^{-2s}{\partial \over \partial z}
\ln \Bigr[z^{-2}I_{1}(z)(I_{1}(z)+zI_{1}'(z))\Bigr] .
\end{equation}
This yields (see the Appendix)
\begin{equation}
{\widetilde \zeta}'(0)=\ln(\pi a^{2}) .
\end{equation}
By virtue of (1.12), (2.9) and (2.11) one finds
\begin{equation}
\zeta_{LN}'(0)=-{631\over 15120}-{13\over 90}\ln(2)
+{1\over 2}\ln(\pi)+{37\over 45}\ln(a^{2})
+{2\over 3}\zeta_{R}'(-3)-{5\over 3}\zeta_{R}'(-1) .
\end{equation}
Bearing in mind that [19] $\zeta_{LN}(0)={37\over 45}$, which
is indeed encoded in (2.12), one obtains
\begin{equation}
\Gamma_{LN}^{(1)}=-{37\over 90}\ln(\mu^{2}a^{2})
+{13\over 180}\ln(2)-{1\over 4}\ln(\pi)
+{631\over 30240}-{1\over 3}\zeta_{R}'(-3)
+{5\over 6}\zeta_{R}'(-1) .
\end{equation}

Last, the contribution of the decoupled normal mode $R_{1}$
should be considered. In the Lorentz gauge, $R_{1}$ takes the
form [19] 
\begin{equation}
R_{1}(\tau)={1\over \tau}I_{2}(M\tau) ,
\end{equation}
up to an unessential multiplicative constant, and hence it
contributes (see the Appendix)
\begin{equation}
\Gamma_{R_{1}}^{(1)}={3\over 8}\ln(\mu^{2}a^{2})
-{1\over 4}\ln(2)+{1\over 4}\ln(\pi) .
\end{equation}
One can now combine Eqs. (2.4), (2.5), (2.13) and (2.15) to
find the full one-loop effective action in the Lorentz gauge as
\begin{equation}
\Gamma^{(1)}={31\over 180}\ln(\mu^{2}a^{2})
+{7\over 45}\ln(2)+{1\over 2}\ln(\pi)
+{6931\over 30240}-{1\over 3}\zeta_{R}'(-3)
+{11\over 6}\zeta_{R}'(-1) .
\end{equation}
Interestingly, this differs both from the result (2.4), which
only involves transverse modes, and from the result for
$\Gamma^{(1)}$ obtained in Ref. [27] in the noncovariant gauge
$\nabla^{\mu}{\cal A}_{\mu}-{2\over 3}{\cal A}_{0}
{\mbox {Tr}}K$:
\begin{equation}
\Gamma_{NC}^{(1)}={77\over 360}\ln(\mu^{2}a^{2})
-{8\over 45}\ln(2)+{1\over 4}\ln(\pi)
+{1991\over 6048}-{1\over 3}\zeta_{R}'(-3)
+{5\over 6}\zeta_{R}'(-1) .
\end{equation}

\section{Two-boundary problem}

Section II has studied a background with boundary which is
more relevant for quantum cosmology (at least in the
Hartle-Hawking program), where one boundary three-surface 
shrinks to zero [1,2]. The standard quantum field theoretical
framework, however, deals with boundary data on {\it two}
boundary three-surfaces, which are necessary to specify
completely the path-integral representation of the propagation
amplitude [34]. Hence we here focus on the one-loop analysis
of Euclidean Maxwell theory in the presence of two concentric
three-sphere boundaries [19].

Since the singularity at the origin ($\tau=0$) is avoided in
this boundary-value problem, the basis functions for the
various modes in (2.1)--(2.3) become linear combinations of
both $I_{n}$ and $K_{n}$ (modified) Bessel functions. We begin
with the most difficult part of the calculation, i.e., the
determinant of the $4 \times 4$ matrix which yields implicitly 
the eigenvalues for longitudinal and normal modes. Such a
matrix, given in Eq. (3.13) of Ref. [19], is obtained by
imposing the boundary conditions (1.1) and (1.4) at the 
three-sphere boundaries of radii $r_{-}$ and $r_{+}$
(hereafter $r_{+}> r_{-}$). Again, one has to express
Bessel functions of various orders in terms of 
$I_{n},I_{n}',K_{n},K_{n}'$ only, where $n \geq 2$. After a
lengthy calculation, such a determinant is found to take the
form
\begin{eqnarray}
{\cal D}&=& {16n^{2}\over (n-1)^{2}M^{2}r_{-}r_{+}}
\Bigr(I_{n}(Mr_{-})K_{n}(Mr_{+})-I_{n}(Mr_{+})K_{n}(Mr_{-})\Bigr)
\nonumber\\
&\times& \biggr \{M^{2}r_{+}r_{-}\Bigr[I_{n}'(Mr_{+})
K_{n}'(Mr_{-})-I_{n}'(Mr_{-})K_{n}'(Mr_{+})\Bigr]
\nonumber\\
&+& Mr_{-}\Bigr[I_{n}(Mr_{+})K_{n}'(Mr_{-})
-I_{n}'(Mr_{-})K_{n}(Mr_{+})\Bigr] \nonumber\\
&+& Mr_{+}\Bigr[I_{n}'(Mr_{+})K_{n}(Mr_{-})
-I_{n}(Mr_{-})K_{n}'(Mr_{+})\Bigr] \nonumber\\
&+& \Bigr[I_{n}(Mr_{+})K_{n}(Mr_{-})
-I_{n}(Mr_{-})K_{n}(Mr_{+})\Bigr] \biggr \} .
\end{eqnarray}
Thus, one has first to multiply (3.1) by $M^{2}$ to get
rid of fake roots [35]. By virtue of the uniform asymptotic
expansions (A1)--(A4), only the effects of $K_{n}(Mr_{-})$
and $I_{n}(Mr_{+})$ survive at large $M$ [19]. Thus, after 
setting $Mr_{+}=zn$ (cf. Sec. II), which implies that
$Mr_{-}=znr_{-}/r_{+}$, the contributions of the $A_{i}$
functions to $\zeta'(0)$ (see (1.5)) can be obtained by
studying
\begin{eqnarray}
\ln{\cal D}&\sim& \ln \Bigr[z^{-n}I_{n}(zn)\Bigr]
+\ln \Bigr[z^{n}K_{n}(znr_{-}/r_{+})\Bigr] \nonumber\\
&+& \ln \Bigr[I_{n}(zn)+znI_{n}'(zn)\Bigr] \nonumber\\
&+& \ln \Bigr[K_{n}(znr_{-}/r_{+})
+zn{r_{-}\over r_{+}}K_{n}'(znr_{-}/r_{+})\Bigr] .
\end{eqnarray}
This means that the asymptotic terms are a sum of Dirichlet
contributions for the inner and outer space, and Robin 
contributions for the inner and outer space with $u=1$. Looking
at the asymptotics of $K_{n}$ and $K_{n}'$, the relation
between inner space and outer space is immediate:
$(A_{-1})_{I}=-(A_{-1})_{K}$,
$(A_{0})_{I}=(A_{0})_{K}$,
$(A_{i})_{I}=(-1)^{i}(A_{i})_{K}$ (see Eqs. (A12)--(A15)). The
resulting asymptotics of (3.2) reads
\begin{equation}
A_{-1}(s)={(r_{+}^{2s}-r_{-}^{2s})\over 2\sqrt{\pi}}
{\Gamma(s-{1\over 2})\over \Gamma(s+1)}\zeta_{H}(2s-3;2) ,
\end{equation}
\begin{equation}
A_{0}(s)=0 ,
\end{equation}
\begin{equation}
A_{i}(s)=-{1\over \Gamma(s)}\Bigr(r_{+}^{2s}-r_{-}^{2s}\Bigr)
\zeta_{H}(2s+i-2;2)\sum_{l=0}^{i}(x_{i,l}+z_{i,l})
{\Gamma(s+l+{i\over 2})\over \Gamma(l+{i\over 2})}, \;
i=1,3 ,
\end{equation}
\begin{equation}
A_{2}(s)=-{1\over \Gamma(s)}\Bigr(r_{+}^{2s}+r_{-}^{2s}\Bigr)
\zeta_{H}(2s;2)\sum_{l=0}^{2}(x_{2,l}+z_{2,l})
{\Gamma(s+l+1)\over \Gamma(l+1)} .
\end{equation}
Note that the second argument of the Hurwitz $\zeta$-function
is 2, to take into account that the infinite sums defining
$A_{i}$ start from $n=2$ in our problem. Moreover, the
$x_{i,l}$ and $z_{i,l}$ are the coefficients of the polynomials
$D_{i}$ (Dirichlet case) and $M_{i}$ (Robin case), 
respectively (see Eqs. (A6)--(A11)). In Eqs. (3.3)--(3.6)
one has now to pick out the coefficients of the terms linear 
in $s$, since these are the only ones which contribute to
$\zeta'(0)$. Hence one finds
\begin{equation}
A_{-1}'(0)={119\over 60}\ln(r_{+}/r_{-}) ,
\end{equation}
\begin{equation}
A_{0}'(0)=A_{1}'(0)=0 ,
\end{equation}
\begin{equation}
A_{2}'(0)=-{3\over 2} ,
\end{equation}
\begin{equation}
A_{3}'(0)=-{61\over 180}\ln(r_{+}/r_{-}) ,
\end{equation} 
which imply
\begin{equation}
\sum_{i=-1}^{3}A_{i}'(0)=-{3\over 2}
+{74\over 45}\ln(r_{+}/r_{-}) .
\end{equation}

The contribution of $Z(s) \equiv \sum_{n=2}^{\infty}
n^{2}Z_{n}(s)$ to $\zeta'(0)$ (cf. (1.6)) involves the
logarithm of the Bessel terms in (3.1), and a further 
contribution resulting from the uniform asymptotics of such
Bessel functions. Hence it reads (see the Appendix for
details)
\begin{eqnarray}
Z'(0)&=& -\sum_{n=2}^{\infty}n^{2} \biggr \{
\ln \Bigr[-I_{n}(Mr_{-})K_{n}(Mr_{+})
+I_{n}(Mr_{+})K_{n}(Mr_{-})\Bigr] \nonumber\\
&+& \ln \Bigr[I_{n}(Mr_{+})K_{n}(Mr_{-})
-I_{n}(Mr_{-})K_{n}(Mr_{+}) \nonumber\\
&+& Mr_{-}\Bigr[I_{n}(Mr_{+})K_{n}'(Mr_{-})
-I_{n}'(Mr_{-})K_{n}(Mr_{+})\Bigr] \nonumber\\
&+& Mr_{+}\Bigr[I_{n}'(Mr_{+})K_{n}(Mr_{-})
-I_{n}(Mr_{-})K_{n}'(Mr_{+})\Bigr] \nonumber\\
&+& M^{2}r_{+}r_{-}\Bigr[I_{n}'(Mr_{+})
K_{n}'(Mr_{-})-I_{n}'(Mr_{-})K_{n}'(Mr_{+})
\Bigr]\Bigr] \nonumber\\
&-& 2n \Bigr[\eta(Mr_{+})-\eta(Mr_{-})\Bigr]
+{1\over n^{2}} \biggr \} ,
\end{eqnarray}
where all Bessel functions should be studied in the limit
as $M \rightarrow 0$ [23]. One can thus use Eqs. (A20) and 
(A21) which express the limiting behavior of Bessel
functions in such a case. Many terms are then found to cancel
each other exactly, leading to
\begin{equation}
Z'(0)=-\sum_{n=2}^{\infty}n^{2} \biggr \{
2 \ln \Bigr[1-(r_{-}/r_{+})^{2n} \Bigr]
+ \ln \Bigr(1-{1\over n^{2}}\Bigr)
+{1\over n^{2}} \biggr \} ,
\end{equation}
where one has [27]
\begin{equation}
-\sum_{n=2}^{\infty}n^{2}\biggr[\ln \Bigr(1-{1\over n^{2}}\Bigr)
+{1\over n^{2}} \biggr]={3\over 2} - \ln(\pi) .
\end{equation}
Equations (3.11), (3.13) and (3.14) imply that
\begin{equation}
\zeta_{LN}'(0)={74\over 45}\ln(r_{+}/r_{-})-\ln(\pi)
-2\sum_{n=2}^{\infty}n^{2}
\ln \Bigr[1-(r_{-}/r_{+})^{2n}\Bigr] .
\end{equation}

Let us now study the decoupled normal mode $R_{1}$. In our
two-boundary problem, it takes the form
\begin{equation}
R_{1}(\tau)={\beta_{1}\over \tau}I_{2}(M\tau)
+{\beta_{2}\over \tau}K_{2}(M\tau) ,
\end{equation}
where $\beta_{1}$ and $\beta_{2}$ are some constants.
By virtue of the boundary condition (1.4), one should set to
zero at the three-sphere boundaries the linear combination
${dR_{1}\over d\tau}+{3\over \tau}R_{1}$. The resulting eigenvalue
condition has no fake roots, and hence one finds (see Sec. IV)
\begin{equation}
\Gamma_{R_{1}}^{(1)}=-{1\over 4}\ln(\mu^{2}r_{+}r_{-})
+{1\over 2}\ln(r_{+}/r_{-})
+{1\over 2}\ln \Bigr[1-(r_{-}/r_{+})^{2}\Bigr] .
\end{equation}

Last, we study the determinants of the $2 \times 2$ matrices which
yield implicitly the eigenvalues for transverse and ghost modes.
In both cases, the eigenvalue condition is
\begin{equation}
I_{n}(Mr_{-})K_{n}(Mr_{+})-I_{n}(Mr_{+})K_{n}(Mr_{-})=0,
\end{equation}
where for ghost modes the integer $n$ is $\geq 1$ and the 
degeneracy is $-2n^{2}$ [19], and for transverse modes the
integer $n$ is $\geq 2$, with degeneracy $2(n^{2}-1)$ [31].
Bearing in mind the ghost degeneracy and Eqs. (1.7)--(1.11), 
the asymptotic contribution for ghosts is expressed by
\begin{equation}
A_{-1}^{\mbox{gh}}(s)=-{1\over 2\sqrt{\pi}}
{\Gamma(s-{1\over 2})\over \Gamma(s+1)}
\Bigr[r_{+}^{2s}-r_{-}^{2s}\Bigr]\zeta_{R}(2s-3),
\end{equation}
\begin{equation}
A_{0}^{\mbox{gh}}(s)={1\over 2}\Bigr[r_{+}^{2s}
+r_{-}^{2s}\Bigr]\zeta_{R}(2s-2),
\end{equation}
\begin{equation}
A_{i}^{\mbox{gh}}(s)={2\over \Gamma(s)}
\Bigr[r_{+}^{2s}+(-1)^{i}r_{-}^{2s}\Bigr]\zeta_{R}(2s+i-2)
\sum_{l=0}^{i}x_{i,l}
{\Gamma(s+l+{i\over 2})\over \Gamma(l+{i\over 2})},
\end{equation}
while for transverse modes one finds
\begin{equation}
A_{-1}^{\mbox{tr}}(s)={1\over 2\sqrt{\pi}}
{\Gamma(s-{1\over 2})\over \Gamma(s+1)}
\Bigr[r_{+}^{2s}-r_{-}^{2s}\Bigr]
\Bigr[\zeta_{R}(2s-3)-\zeta_{R}(2s-1)\Bigr],
\end{equation}
\begin{equation}
A_{0}^{\mbox{tr}}(s)=-{1\over 2}\Bigr[r_{+}^{2s}
+r_{-}^{2s}\Bigr]\Bigr[\zeta_{R}(2s-2)-\zeta_{R}(2s)
\Bigr],
\end{equation}
\begin{equation}
A_{i}^{\mbox{tr}}(s)=-{2\over \Gamma(s)}
\Bigr[r_{+}^{2s}+(-1)^{i}r_{-}^{2s}\Bigr]
\Bigr[\zeta_{R}(2s+i-2)-\zeta_{R}(2s+i)\Bigr]
\sum_{l=0}^{i}x_{i,l}{\Gamma(s+l+{i\over 2})\over
\Gamma(l+{i\over 2})}.
\end{equation}
The most convenient way to proceed is now to add up the
contributions (3.19)--(3.21) and (3.22)--(3.24). This yields 
\begin{equation}
A_{-1}(s)=-{1\over 2\sqrt{\pi}}
{\Gamma(s-{1\over 2})\over \Gamma(s+1)}
\Bigr[r_{+}^{2s}-r_{-}^{2s}\Bigr]\zeta_{R}(2s-1),
\end{equation}
\begin{equation}
A_{0}(s)={1\over 2}\Bigr[r_{+}^{2s}+r_{-}^{2s}\Bigr]
\zeta_{R}(2s),
\end{equation}
\begin{equation}
A_{i}(s)={2\over \Gamma(s)}\Bigr[r_{+}^{2s}
+(-1)^{i}r_{-}^{2s}\Bigr]\zeta_{R}(2s+i)
\sum_{l=0}^{i}x_{i,l}{\Gamma(s+l+{i\over 2})\over
\Gamma(l+{i\over 2})}.
\end{equation}
As in the previous cases, the contributions to $\zeta'(0)$
of (3.25)--(3.27) are obtained by considering their expansion
in the neighboorhood of $s=0$, and adding the coefficients of 
all terms linear in $s$. This yields
\begin{equation}
\sum_{i=-1}^{3}A_{i}'(0)= -{1\over 3}\ln(r_{+}/r_{-})
-{1\over 2}\ln(4\pi^{2}r_{+}r_{-}).
\end{equation}

Moreover, the form of $Z'(0)$ for ghost and transverse modes
is considerably simplified because the full degeneracy 
is $-2$. This leads to (cf. Eqs. (3.12) and (3.13))
\begin{eqnarray}
{Z^{\mbox{gh},\mbox{tr}}}'(0)&=& 
2\sum_{n=1}^{\infty} \biggr \{
\ln \Bigr[-I_{n}(Mr_{-})K_{n}(Mr_{+})
+I_{n}(Mr_{+})K_{n}(Mr_{-})\Bigr] \nonumber \\
&-&n \Bigr[\eta(Mr_{+})-\eta(Mr_{-})\Bigr]
+\ln(-2n) \biggr \} \nonumber\\
&=& 2\sum_{n=1}^{\infty}\ln 
\Bigr[1-(r_{-}/r_{+})^{2n}\Bigr].
\end{eqnarray}
Last, since ghost modes yield a vanishing contribution to
$\zeta(0)$, while transverse modes contribute $-{1\over 2}$,
one finds
\begin{equation}
\Gamma^{(1)}_{\mbox{gh},\mbox{tr}}
={1\over 4}\ln(\mu^{2}r_{+}r_{-})
-\sum_{n=1}^{\infty}\ln \Bigr[1-(r_{-}/r_{+})^{2n}\Bigr]
+{1\over 6}\ln(r_{+}/r_{-})
+{1\over 2}\ln(2\pi).
\end{equation}
The results (3.15), (3.17) and (3.30) lead to the following
value of the one-loop effective action in the two-boundary
problem:
\begin{eqnarray} 
\Gamma^{(1)}&=&-{7\over 45}\ln(r_{+}/r_{-})
+{1\over 2}\ln(2)+\ln(\pi)
-{1\over 2}\ln \Bigr[1-(r_{-}/r_{+})^{2}\Bigr] \nonumber \\
&+&\sum_{n=1}^{\infty}(n^{2}-1)\ln \Bigr[1-(r_{-}/r_{+})^{2n}
\Bigr].
\end{eqnarray}

\section{Barvinsky-Kamenshchik-Karmazin-Mishakov Technique}

When a series of difficult calculations is performed, it is
appropriate to have an independent check of the final result.
For this purpose, we here outline the application of the 
technique described in Refs. [14,15,35]. Let $f_{n}$ be the
function occurring in the equation obeyed by the eigenvalues 
by virtue of boundary conditions, and let $d(n)$ be the degeneracy
of such eigenvalues labelled by the integer $n$. One then
defines the function
\begin{equation}
I(M^{2},s) \equiv \sum_{n=n_{0}}^{\infty}d(n)n^{-2s}
\ln f_{n}(M^{2}) ,
\end{equation}
where $M^{2}$ is related to the eigenvalues through the relation
$M^{2}=-\lambda_{n}$, and fake roots (e.g., $M=0$ for Bessel
functions) have been taken out when $f_{n}$ is written down. 
The function (4.1) admits an analytic continuation to the
complex-$s$ plane as a meromorphic function with a simple pole
at $s=0$: i.e.,
\begin{equation}
``I(M^{2},s)"={I_{\mbox {pole}}(M^{2})\over s}
+I^{R}(M^{2})+{\mbox O}(s).
\end{equation}
The functions occurring on the right-hand side of (4.2) make it
possible to evaluate $\zeta(0)$ and $\zeta'(0)$ in quantum field
theory as
\begin{equation}
\zeta(0)=I_{\mbox {log}}+I_{\mbox {pole}}(M^{2}=\infty)
-I_{\mbox {pole}}(M^{2}=0),
\end{equation}
\begin{equation}
\zeta'(0)=I^{R}(M^{2}=\infty)-I^{R}(M^{2}=0)
-\int_{0}^{\infty}\ln(M^{2}){dI_{\mbox {pole}}(M^{2})
\over dM^{2}} dM^{2},
\end{equation}
where $I_{\mbox {log}}$ is the coefficient of $\ln(M)$ in the
uniform asymptotic expansion of $I(M^{2},s)$, after taking out
fake roots. In quantum mechanics, as well as for decoupled modes
of a quantum field, $\zeta(0)$ reduces to $I_{\mbox {log}}$,
and $\zeta'(0)$ reduces to $I^{R}(\infty)-I^{R}(0)$. 

In Eq. (3.16), the decoupled mode gives rise to the eigenvalue 
condition
\begin{eqnarray}
0&=& \left(I_{2}'(Mr_{+})+2{I_{2}(Mr_{+})\over Mr_{+}}\right)
\left(K_{2}'(Mr_{-})+2{K_{2}(Mr_{-})\over Mr_{-}}\right)
\nonumber \\
&-& \left(I_{2}'(Mr_{-})+2{I_{2}(Mr_{-})\over Mr_{-}}\right)
\left(K_{2}'(Mr_{+})+2{K_{2}(Mr_{+})\over Mr_{+}}\right).
\end{eqnarray}
Thus, Eqs. (4.1), (4.2) and (A1)--(A4) lead to
\begin{equation}
I^{R}(\infty)=-\ln(2)-{1\over 2}\ln(r_{+}r_{-}).
\end{equation}
Moreover, $I^{R}(0)$ is obtained from the limiting behavior
of Bessel functions as $M \rightarrow 0$ (see (A20) and (A21)),
and one finds
\begin{equation}
I^{R}(0)=-\ln(2)+\ln(r_{+}/r_{-})
+\ln \Bigr(1-(r_{-}/r_{+})^{2}\Bigr) .
\end{equation}
The result (3.17) is obtained if one bears in mind that the
decoupled mode provides an example of a nontrivial eigenfunction
obeying the boundary conditions and belonging to the zero
eigenvalue. Hence one deals with a one-dimensional null space
whose dimension should be added to $I_{\mbox{log}}$ to obtain
the correct $\zeta(0)$ value for $R_{1}$ as $\zeta(0)
={1\over 2}$ [19].

It is also instructive to outline the ghost calculation in the
two-boundary problem. The function of Eq. (4.1) takes then
the form
\begin{equation}
I(M^{2},s)=2\sum_{n=1}^{\infty}n^{2-2s}\ln \Bigr[
I_{n}(nMr_{+})K_{n}(nMr_{-})-I_{n}(nMr_{-})K_{n}(nMr_{+})
\Bigr].
\end{equation}
As $M \rightarrow \infty$, only $K_{n}(nMr_{-})$ and
$I_{n}(nMr_{+})$ contribute to $\zeta'(0)$, and the uniform
asymptotic expansions (A1) and (A3) imply that
\begin{equation}
I^{R}(\infty)=0,
\end{equation}
since $\zeta_{R}(-2)=0$. To evaluate $I^{R}(0)$ one needs
instead the limiting behavior of Bessel functions as
$M \rightarrow 0$. Thus, the expansions (A20) and (A21)
lead to
\begin{equation}
I^{R}(0)=-2\sum_{n=1}^{\infty}n^{2}\ln(n)+{1\over 60}
\ln(r_{+}/r_{-})+2\sum_{n=1}^{\infty}n^{2}
\ln \Bigr[1-(r_{-}/r_{+})^{2n}\Bigr].
\end{equation}
Last, the third term on the right-hand side of (4.4) is
found to be
\begin{equation}
-\int_{0}^{\infty}\ln(M^{2}){dI_{\mbox {pole}}
\over dM^{2}} dM^{2}
=-{1\over 180}\ln(r_{+}/r_{-}).
\end{equation}
Equations (4.9)--(4.11) yield
\begin{equation}
\zeta'(0)_{\mbox{gh}}=-{1\over 45}\ln(r_{+}/r_{-})
-2\sum_{n=1}^{\infty}n^{2}\ln \Bigr[1-(r_{-}/r_{+})^{2n}
\Bigr] -2\zeta_{R}'(-2) ,
\end{equation}
and this should be multiplied by $-1$, since the ghost
contribution has fermionic nature for a bosonic field.

For transverse modes, an analogous procedure yields
\begin{eqnarray}
\zeta'(0)_{\mbox{tr}}&=&-{16\over 45}\ln(r_{+}/r_{-})
-{1\over 2}\ln(r_{+}r_{-})
-2\sum_{n=1}^{\infty}(n^{2}-1)\ln
\Bigr[1-(r_{-}/r_{+})^{2n}\Bigr] \nonumber \\
&-&\ln(2\pi) -2 \zeta_{R}'(-2).
\end{eqnarray}
Equations (4.12) and (4.13) yield a result in complete
agreement with Eq. (3.30).

Last, to evaluate the contribution of longitudinal and
normal modes one starts from the determinant (3.1),
which implies
\begin{equation}
I^{R}(\infty)=\sum_{n=2}^{\infty}n^{2}
\ln {n^{2}\over (n-1)^{2}r_{+}r_{-}},
\end{equation}
\begin{equation}
I^{R}(0)=\sum_{n=2}^{\infty}{1\over r_{+}r_{-}}
\Bigr[(r_{+}/r_{-})^{2n}-(r_{-}/r_{+})^{2n}\Bigr]^{2}
{(n^{2}-1)\over (n-1)^{2}}.
\end{equation}
Moreover, the third term on the right-hand side of (4.4)
contributes
\begin{equation}
-\int_{0}^{\infty}\ln(M^{2}){dI_{\mbox{pole}}
\over dM^{2}}dM^{2}=-{61\over 180}\log(r_{+}/r_{-}).
\end{equation}
The results (4.14)--(4.16), jointly with (3.14), lead to
$\zeta_{LN}'(0)$ as in (3.15).
   
\section{Results and open problems}

Our paper has studied in detail the one-loop effective
action $\Gamma^{(1)}$ for Euclidean Maxwell theory on
manifolds with boundary, in the case of flat four-space
bounded by a three-sphere, or two concentric three-spheres.
Our main result is that, by using covariant gauges such as
the Lorentz gauge within the framework of Faddeev-Popov
formalism for semiclassical amplitudes, $\Gamma^{(1)}$ (see
(2.16)) differs from the contribution of transverse modes obtained
in Refs. [25,26] (see (2.4)), and it also differs from the value
obtained in Ref. [27] in the noncovariant gauge
$\nabla^{\mu}{\cal A}_{\mu}-{2\over 3}{\cal A}_{0}
{\mbox {Tr}}K$ (see (2.17)). This is confirmed by the two-boundary
analysis of Sec. III (see Eq. (3.31)). Our result 
seems to add evidence in favor of longitudinal, normal and ghost
modes not being able to cancel each other's effects exactly on
manifolds with boundary [17--20].

The extension to curved backgrounds (e.g., $S^{4}$ bounded by
$S^{3}$) is of purely technical nature and can be obtained after
dealing properly with the asymptotics of Legendre functions (instead
of the Bessel functions occurring in the flat case). At least
three outstanding problems, however, remain. First, one would like
to evaluate the one-loop effective action for the family of 
noncovariant gauges $\nabla^{\mu}{\cal A}_{\mu}-b{\cal A}_{0}
{\mbox {Tr}}K$, first introduced in Ref. [17], and then studied
extensively in Refs. [19,20]. These gauges have been criticized in
Ref. [36], but they appear a necessary step to complete the
quantization program in arbitrary gauges, at least in the
semiclassical approximation.

Second, one would like to apply our algorithms to the analysis
of Euclidean quantum gravity. For this purpose, one can impose
the boundary conditions in terms of (complementary) projectors
proposed in Ref. [5], or the boundary conditions completely
invariant under infinitesimal diffeomorphisms [4], or the boundary
conditions of Ref. [8], which are Robin on $h_{ij}$ and the
ghost one-form, and Dirichlet on normal components $h_{00}$
and $h_{0i}$ of metric perturbations. Yet another possibility
is represented by the nonlocal boundary conditions of 
Ref. [9]. In such cases, it would be
interesting to investigate the asymptotics of the eigenvalue
condition given by the vanishing of the determinant of an
$8 \times 8$ matrix in the two-boundary problem for pure
gravity. More work can be done in this respect.

Last, but not least, geometric formulas for $\zeta'(0)$ are 
still lacking in arbitrary gauges on manifolds with boundary.
What happens is that the usual Schwinger-DeWitt method
fails to hold for nonminimal operators resulting from the
choice of arbitrary gauge-averaging terms in the Euclidean
action. More precisely, the factor which stands before the 
series in $t$ in the heat-kernel asymptotics is not a
Gaussian but a complicated special function. This leads in
turn to relations for heat-kernel coefficients unbounded 
from below as well as above, and hence these equations cannot
be solved recursively [37].
Nevertheless, if one were able to generalize the
technique developed in Refs. [38,39] to manifolds with boundary,
one would obtain an independent check of the several analytic
results which can be derived in the near future. This would
lead in turn to a much deeper understanding of heat-kernel
asymptotics for quantized gauge fields and quantum gravity 
on manifolds with boundary.

Such an investigation is regarded by the authors as an
important task for the years to come, and it makes us feel
that a new exciting age is in sight in the application of
heat-kernel methods to quantum field theory.

\acknowledgements
We are grateful to many colleagues for correspondence and/or
scientificic collaboration on the one-loop effective action.
In particular, we should mention 
Ivan Avramidi, Michael Bordag, Stuart Dowker,
Emilio Elizalde and Giuseppe Pollifrone.
The work of A.Y.K. was partially supported by the Russian
foundation for Fundamental Researches through Grant No.
96-02-16220-a, and by the Russian Research Project 
``Cosmomicrophysics". K.K. acknowledges financial support
from the DFG, contract Bo 1112/4-1.

\appendix
\section*{}

The uniform asymptotic expansions as $\rho \rightarrow 
\infty$ of the Bessel functions $I_{\rho}(\rho z),
K_{\rho}(\rho z)$, jointly with their first derivatives, are
derived in detail in Ref. [40], and they play a fundamental
role in the analytic investigation of conformal anomalies
and one-loop effective action. In terms of the Debye polynomials
$u_{k}(t)$ and $v_{k}(t)$ [41], they read
\begin{equation}
I_{\rho}(\rho z) \sim {1\over \sqrt{2\pi \rho}}
{e^{\rho \eta}\over (1+z^{2})^{1\over 4}}
\left[1+\sum_{k=1}^{\infty}{u_{k}(t)\over \rho^{k}}\right] ,
\end{equation}
\begin{equation}
I_{\rho}'(\rho z) \sim {1\over \sqrt{2\pi \rho}}e^{\rho \eta}
{(1+z^{2})^{1\over 4}\over z}\left[1+\sum_{k=1}^{\infty}
{v_{k}(t)\over \rho^{k}}\right] ,
\end{equation}
\begin{equation}
K_{\rho}(\rho z) \sim \sqrt{\pi \over 2\rho}
{e^{-\rho \eta}\over (1+z^{2})^{1\over 4}}
\left[1+\sum_{k=1}^{\infty}(-1)^{k}{u_{k}(t)\over 
\rho^{k}}\right] ,
\end{equation}
\begin{equation}
K_{\rho}'(\rho z) \sim -\sqrt{\pi \over 2\rho}
e^{-\rho \eta}{(1+z^{2})^{1\over 4}\over z}
\left[1+\sum_{k=1}^{\infty}(-1)^{k}{v_{k}(t)\over \rho^{k}}
\right] ,
\end{equation}
where $t \equiv {1\over \sqrt{1+z^{2}}}$, and
$\eta \equiv \sqrt{1+z^{2}}+\ln [z/(1+\sqrt{1+z^{2}})]$. In
the one-loop analysis it is necessary to evaluate the 
logarithm of the equation obeyed by the eigenvalues by virtue
of boundary conditions. In particular, we need the asymptotic
expansion [27]
\begin{equation}
\ln \left[1+\sum_{k=1}^{\infty}{u_{k}(t)\over \rho^{k}}
\right] \sim \sum_{p=1}^{\infty}{D_{p}(t)\over \rho^{p}} ,
\end{equation}
where [27]
\begin{equation}
D_{1}(t)={1\over 8}t-{5\over 24}t^{3} ,
\end{equation}
\begin{equation}
D_{2}(t)={1\over 16}t^{2}-{3\over 8}t^{4}
+{5\over 16}t^{6} ,
\end{equation}
\begin{equation}
D_{3}(t)={25\over 384}t^{3}-{531\over 640}t^{5}
+{221\over 128}t^{7}-{1105\over 1152}t^{9} .
\end{equation}
In the case of Robin boundary conditions, a linear combination
of $I_{\rho}$ and $I_{\rho}'$ is set to zero at the boundary, 
and a dimensionless parameter $u$ occurs in the eigenvalue
condition. Thus, the polynomials (A6)--(A8) are replaced by [27]
\begin{equation}
M_{1}(t,u)=\Bigr(-{3\over 8}+u \Bigr)t+{7\over 24}t^{3} ,
\end{equation}
\begin{equation}
M_{2}(t,u)=\Bigr(-{3\over 16}+{1\over 2}u
-{1\over 2}u^{2}\Bigr)t^{2}
+\Bigr({5\over 8}-{1\over 2}u \Bigr)t^{4}-{7\over 16}t^{6} ,
\end{equation}
\begin{eqnarray}
M_{3}(t,u)&=& \Bigr(-{21\over 128}+{3\over 8}u-{1\over 2}u^{2}
+{1\over 3}u^{3}\Bigr)t^{3}
+ \Bigr({869\over 640}-{5\over 4}u
+{1\over 2}u^{2}\Bigr)t^{5}
\nonumber\\
&+& \Bigr(-{315\over 128}+{7\over 8}u \Bigr)t^{7}
+{1463\over 1152}t^{9} .
\end{eqnarray}
When also $K_{\rho}$ functions occur in the calculation of
functional determinants, one has polynomials 
${\widetilde D}_{i}(t)=(-1)^{i}D_{i}(t)$, and
${\widetilde M}_{i}(t,u)=(-1)^{i}M_{i}(t,u)$.

In the case of Dirichlet boundary conditions, the functions
(1.7)--(1.11) are infinite sums of the contributions 
[23,27,28]
\begin{equation}
A_{-1}^{l}={\sin(\pi s)\over \pi}\int_{0}^{\infty}
dz (zl/a)^{-2s}{\partial \over \partial z}
\ln \left({z^{-l}\over \sqrt{2\pi l}}e^{l\eta}\right) ,
\end{equation}
\begin{equation}
A_{0}^{l}={\sin(\pi s)\over \pi}\int_{0}^{\infty}
dz (zl/a)^{-2s}{\partial \over \partial z}
\ln (1+z^{2})^{-{1\over 4}} ,
\end{equation}
\begin{equation}
A_{i}^{l}={\sin(\pi s)\over \pi}\int_{0}^{\infty}
dz (zl/a)^{-2s}{\partial \over \partial z}
\left({D_{i}(t)\over l^{i}}\right) .
\end{equation}
In the two-boundary problems, however, also $K$ functions
and their first derivatives contribute. By virtue of (A3),
(A4), (A12)--(A14) one thus finds
\begin{equation}
(A_{-1})_{I}=-(A_{-1})_{K} \; , \;
(A_{0})_{I}=(A_{0})_{K} \; , \;
(A_{i})_{I}=(-1)^{i}(A_{i})_{K} ,
\end{equation}
for both Dirichlet and Robin boundary conditions. This
leads to (3.3)--(3.6).

The recurrence relations among Bessel functions used in the
course of deriving (2.7) from (2.6) are
\begin{equation}
I_{n-2}(z)=\Bigr(1+{2n(n-1)\over z^{2}}\Bigr)I_{n}(z)
+2{(n-1)\over z}I_{n}'(z) ,
\end{equation}
\begin{equation}
I_{n-1}(z)=I_{n}'(z)+{n\over z}I_{n}(z) ,
\end{equation}
\begin{equation}
I_{n+1}(z)=I_{n}'(z)-{n\over z}I_{n}(z) ,
\end{equation}
\begin{equation}
I_{n+2}(z)=\Bigr(1+{2n(n+1)\over z^{2}}\Bigr)
I_{n}(z)-2{(n+1)\over z}I_{n}'(z) .
\end{equation}

The limiting behavior of Bessel functions as
$z \rightarrow 0$, which is necessary to deal properly with
(3.12), is
\begin{equation}
I_{n}(z) \sim {(z/2)^{n}\over \Gamma(n+1)} ,
\end{equation}
\begin{equation}
K_{n}(z) \sim {1\over 2}\sum_{k=0}^{n-1}(-1)^{k}
{(n-k-1)!\over k!(z/2)^{n-2k}}.
\end{equation}

The contribution (2.15) to the one-loop effective action is the
contribution of the decoupled normal mode $R_{1}$, and it
is best tackled in terms of the algorithm of Ref. [35]. The
general structure of $\zeta'(0)$ is then (see Sec. IV)
$\zeta'(0)=I^{R}(\infty)-I^{R}(0)$, where
\begin{equation}
I^{R}(\infty)=-{1\over 2}\ln(a)-{1\over 2}\ln(2)
-{1\over 2}\ln(\pi),
\end{equation}
\begin{equation}
I^{R}(0)=\ln(a)-\ln(2).
\end{equation}
Moreover, $I_{\mbox {log}}=-{3\over 4}$ [19]. 
Hence one gets (2.15).

Last, we should say that the term on the last line of Eq. (3.12)
reads
\begin{equation}
\sigma = -\sum_{n=2}^{\infty}\left[-2n \Bigr(\eta(Mr_{+})
-\eta(Mr_{-})\Bigr)-{F_{2}(1)\over n^{2}}\right] ,
\end{equation}
where
\begin{equation}
-2n \Bigr(\eta(Mr_{+})-\eta(Mr_{-})\Bigr) \sim 
\ln(r_{+}/r_{-})^{-2n} ,
\end{equation}
while, with the notation of Eqs. (A6)--(A11), one has
\begin{equation}
F_{2}(1)=2D_{2}(1)+2M_{2}(1,1)=-1 .
\end{equation}

\end{document}